\documentclass[journal,twocolumn]{IEEEtran}
\usepackage[dvipdfmx]{graphicx}

\usepackage{amsmath}
\usepackage{amssymb}
\usepackage{color}
\DeclareMathAlphabet{\bm}{OML}{cmm}{b}{it}
\newtheorem{theorem}{Theorem}
\newtheorem{lemma}[theorem]{Lemma}
\newtheorem{definition}[theorem]{Definition}
\newtheorem{corollary}[theorem]{Corollary}
\newtheorem{remark}[theorem]{Remark}

\newcommand{\qed}{\hfill \IEEEQED}

\newcommand{\markov}{\leftrightarrow}

\newcommand{\bol}[1]{\mathbf{#1}}

\allowdisplaybreaks[2]

\begin{document}

\title{Cognitive Interference Channels with Confidential Messages under Randomness Constraint}

\author{Shun~Watanabe~\IEEEmembership{Member,~IEEE}       
\thanks{The first author is with the Department
of Information Science and Intelligent Systems, 
University of Tokushima,
2-1, Minami-josanjima, Tokushima,
770-8506, Japan, 
e-mail:shun-wata@is.tokushima-u.ac.jp.}
and Yasutada~Oohama~\IEEEmembership{Member,~IEEE}       
\thanks{The second author is with the Department of Communication Engineering and Informatics,
University of Electro-Communications, Tokyo, 182-8585, Japan, e-mail:oohama@uec.ac.jp.}

\thanks{Manuscript received ; revised }}

\markboth{Journal of \LaTeX\ Class Files,~Vol.~6, No.~1, January~2007}%
{Shell \MakeLowercase{\textit{et al.}}: Bare Demo of IEEEtran.cls for Journals}

\maketitle
\begin{abstract}
The cognitive interference channel with confidential messages (CICC)
proposed by Liang {\em et.~al.} is investigated.
When the security is considered in coding systems, it is well known
that the sender needs to use a stochastic encoding to avoid the
information about the transmitted confidential message to be
leaked to an eavesdropper. For the CICC, the trade-off between the
rate of the random number to realize the stochastic encoding and the
communication rates is investigated, and the optimal trade-off is completely
characterized.
\end{abstract}


\IEEEpeerreviewmaketitle

\section{Introduction}

\begin{figure}[t]
\centering
\includegraphics[width=\linewidth]{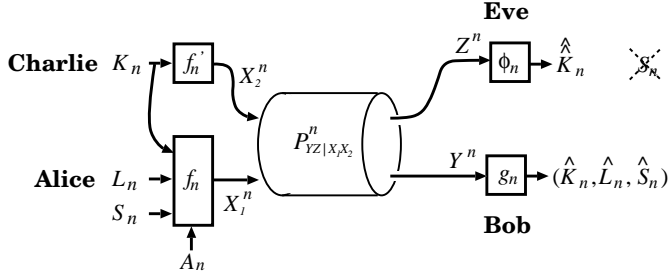}
\caption{The coding system investigated in this paper. Alice sends common
message $K_n$, private message $L_n$, and confidential message $S_n$
by using a deterministic function $f_n$ and a limited amount of dummy randomness $A_n$.
Charlie also sends a signal $X_2^n$ which is a deterministic function of the common
message $K_n$.
The common message is supposed to be decoded by both Bob and Eve.
The private message is supposed to be decoded by Bob, and we do not 
care whether Eve can decode the private message or not.
The confidential message is supposed to be decoded by Bob, and it
must be kept completely secret from Eve.
}
\label{Fig:system}
\end{figure}

Cognitive radio has attracted considerable attention recently,
for it can improve the spectrum efficiency of wireless networks
\cite{mitola-dthesis}. In information theoretical study of the cognitive radio, 
it is usually modeled by a interference channel called cognitive interference channel (CIC),
in which the cognitive transmitter can non-causally know the other transmitter's message
\cite{devroye:06, devroye:06b,wu:07, jovicic:09}. 
We consider the (CIC) model investigated by  Jiang {\em et.~al.} \cite{jiang:08},
Zhong {\em et.~al.} \cite{zhong:07}, and Liang {\em et.~al.} \cite{liang:09},
in which one receiver needs to decode both messages.
Especially as in \cite{liang:09}, we also consider the security,
i.e., the message sent by the cognitive transmitter must be kept secret
from one of the receivers. We call this problem the cognitive
interference channel with confidential messages (CICC).
The coding system investigated in this paper
is described in Fig.~\ref{Fig:system}.

When the security is considered, it is well known that
the sender needs to use a stochastic encoder to
avoid the information about the transmitted confidential
message to be leaked to the eavesdropper Eve.
The stochastic encoder is usually realized by preparing
a dummy random number in addition to the intended messages
and by encoding them to a transmitted signal by a deterministic encoder.
Furthermore, random numbers are also needed to realize the coding
technique called channel prefixing.

In literatures of information theoretic security (eg.~\cite{wyner:75, csiszar:78, liang-book}),
the random number has been regarded as free resource, and the amount of the random
number used in the stochastic encoding has been paid no attention.
However in practice, the random number is quite precious resource.
For example, generation rates of any existing true random number generators
are not as fast as communication rates of wireless networks \cite{Hardware-RNG}.
Although the random number generator equipped in the forthcoming
Intel's CPU can generate the random number as fast as $3$Gbps \cite{IvyBridge},
the communication rate of  the new IEEE wireless communication standard is said to
be over Gbps \cite{802.11ac}. Thus, the random number should be 
regarded as at least as precious as communication resources.
For this purpose, we formulate the problem of the CICC by
randomness constrained stochastic encoder, and completely
characterize the capacity region of this new problem.
We assume that the non-cognitive transmitter, Charlie, only uses
a deterministic encoding. This assumption seems natural because
Charlie only observes the common message, and the common
message need not to be kept secret.
 
The present problem to consider the CICC by the randomness constrained 
stochastic encoder is an extension of the authors' series of works.
In \cite{oohama:10}, the authors investigated the capacity region of
the relay channel with confidential messages for the completely
deterministic encoder, and the capacity region
of the broadcast channel with confidential messages (BCC) for the 
completely deterministic encoder was characterized as a corollary.
In \cite{watanabe:12}, the authors completely characterized the
capacity region of the BCC by the randomness constrained 
stochastic encoder. The problem formulation in this paper is
the extension of that in \cite{watanabe:12} to the CIC, and more
involved coding techniques are needed.

Since the security criterion employed in this paper is slightly 
different from that in \cite{liang:09}, it should be remarked.
In \cite{liang:09}, the cognitive transmitter, Alice, sends two kinds
of messages, the common message and the confidential message, and
the level of secrecy of the confidential message
was evaluated by the equivocation rate. In this paper, Alice sends three kinds
of messages, the common message, the private message, and the confidential
message. The role of the common message is the same as that in \cite{liang:09}.
The private message is supposed to be decoded by one of the receiver, Bob, and
we do not care whether Eve can decode the private message or not.
On the other hand, the confidential message is supposed to be decoded by Bob,
and it must be kept completely secret from Eve. The secrecy of the confidential message is
evaluated by the so-called strong security criterion \cite{maurer:94b, csiszar:96}.
As a byproduct, our direct coding theorem is stronger than that in \cite{liang:09},
i.e., our theorem states the strong secrecy.

The  reason we do not use the equivocation rate formulation is as follows.
In the conventional equivocation rate formulation, if the rate of dummy randomness
is not sufficient, a part of the confidential message is sacrificed to make the other
part completely secret and the rate of the completely secret part corresponds
to the equivocation rate. We  think that the rates of sacrificed part and completely secret part
become clearer by employing our formulation.


The rest of this paper is organized as follows.
In Section \ref{section:problem-main}, the problem
formulation is explained and main results are presented.
In Section \ref{section:proof-main}, the proof of the main 
theorem is presented. Some technical arguments are
presented in Appendices.

\section{Problem Formulation and Main Results}
\label{section:problem-main}

Let $P_{Y|X_1 X_2}$ and $P_{Z|X_1 X_2}$ be two channels with common
input alphabets ${\cal X}_1 \times {\cal X}_2$ and output alphabets ${\cal Y}$ and
${\cal Z}$ respectively. Throughout the paper, the alphabets are assumed to
be finite though we do not use finiteness of the alphabet except
cardinality bonds on auxiliary random variables. 

Let ${\cal K}_n$ be the set of the common message, ${\cal L}_n$ be the set of
the private message, and ${\cal S}_n$ be the set of the confidential message.
The common message is supposed to be decoded by both Bob and Eve.
The private message is supposed to be decoded by Bob, and we do not care
whether Eve can decode the private message or not. The confidential message
is supposed to be decoded by Bob, and it must be kept completely secret from Eve.

Typically, Alice use a stochastic encoder to make the confidential message secret
from Eve, and it is practically realized by using a uniform dummy randomness on
the alphabet ${\cal A}_n$. When the size $|{\cal A}_n|$ of dummy randomness is
infinite, any stochastic encoder from ${\cal K}_n \times {\cal L}_n \times {\cal S}_n$ to ${\cal X}_1^n$
can be simulated by a deterministic encoder 
$f_n:{\cal K}_n \times {\cal L}_n \times {\cal S}_n \times {\cal A}_n \to {\cal X}^n$.
But we are interested in the case with bounded size  $|{\cal A}_n|$ in this paper.
In this paper, we assume that Charlie only use a deterministic encoder 
$f^\prime_n:{\cal K}_n \to {\cal X}_2^n$.

Bob's decoder is defined by function $g_n: {\cal Y}^n \to {\cal K}_n \times {\cal L}_n \times {\cal S}_n$
and the error probability is defined as 
\begin{eqnarray}
\lefteqn{ P_{err}(f_n,f_n^\prime,g_n) } \nonumber \\ 
	&=& \sum_{k_n \in {\cal K}_n} \sum_{\ell_n \in {\cal L}_n} \sum_{s_n \in {\cal S}_n}
	\sum_{a_n \in {\cal A}_n} \frac{1}{|{\cal K}_n||{\cal L}_n||{\cal S}_n||{\cal A}_n|} \nonumber \\
	&& P_{Y|X_1 X_2}^n(y^n|f_n(k_n,\ell_n,s_n,a_n), f^\prime_n(k_n)) \nonumber \\
	&& \bol{1}[g_n(y^n) \neq (k_n,\ell_n,s_n)], 
	\label{eq:bob-decoding-error-probability}
\end{eqnarray}
where $\bol{1}[\cdot]$ is the indicator function. 
Eve's decoder is defined by function $\phi_n:{\cal Z}^n \to {\cal K}_n$ and
the error probability $P_{err}(f_n,f_n^\prime,\phi_n)$ 
is defined in a similar manner as Eq.~(\ref{eq:bob-decoding-error-probability}).

Let 
\begin{eqnarray*}
P_{\tilde{Z}^n|S_n}(z^n|s_n)
  &=& \sum_{k_n \in {\cal K}_n} \sum_{\ell_n \in {\cal L}_n} 
  	\sum_{a_n \in {\cal A}_n} \frac{1}{|{\cal K}_n||{\cal L}_n||{\cal A}_n|} \\
 &&~	P_{Z|X_1 X_2}^n(z^n|f_n(k_n,\ell_n,s_n,a_n),f_n^\prime(k_n)), \\
P_{\tilde{Z}^n}(z^n) 
  &=& \sum_{s_n \in {\cal S}_n} 
\frac{1}{|{\cal S}_n|}  
P_{\tilde{Z}^n|S_n}(z^n|s_n)
\end{eqnarray*}
be the output distributions of the channel $P_{Z|X_1 X_2}^n$.
In this paper, we consider the security criterion given by
\begin{eqnarray*}
D(f_n,f_n^\prime) 
&:=& D(P_{S_n \tilde{Z}^n} \| P_{S_n} \times P_{\tilde{Z}^n}) \\
&=& \sum_{s_n \in {\cal S}_n} \frac{1}{|{\cal S}_n|} D(P_{\tilde{Z}^n|S_n}(\cdot|s_n) \| P_{\tilde{Z}^n}) \\
&=& I(S_n; \tilde{Z}^n),
\end{eqnarray*}
where $D(\cdot \| \cdot)$ is the divergence, and $I(\cdot; \cdot)$ is the mutual information
\cite{cover}. The coding system investigate in this paper is depicted in Fig.~\ref{Fig:system}.

In this paper, we are interested in the trade-off among
the rate the dummy randomness, and
the rates of the common, private, and confidential messages.
\begin{definition}
The rate quadruple $(R_d,R_0,R_1, R_s)$ is said to be {\em achievable} if
there exists a sequence of Alice's deterministic encoder 
$f_n:{\cal K}_n \times {\cal L}_n \times {\cal S}_n \times {\cal A}_n \to {\cal X}_1^n$, 
Charlie's deterministic encoder $f_n^\prime: {\cal K}_n \to {\cal X}_2^n$,
Bob's decoder $g_n:{\cal Y}^n \to {\cal K}_n \times {\cal L}_n \times {\cal S}_n$,
and Eve's decoder $\phi_n:{\cal Z}^n \to {\cal K}_n$ such that
\begin{eqnarray}
\lim_{n \to \infty} P_{err}(f_n,f_n^\prime,g_n) &=& 0, \\
\lim_{n \to \infty} P_{err}(f_n,f_n^\prime,\phi_n) &=& 0, \\
\lim_{ n \to \infty} D(f_n,f_n^\prime) &=& 0, \label{eq:definition-of-achievability-D} \\
\limsup_{n \to \infty} \frac{1}{n} \log |{\cal A}_n| &\le& R_d, \\
\liminf_{n \to \infty} \frac{1}{n} \log |{\cal K}_n| &\ge& R_0, \\
\lim_{n \to \infty} \frac{1}{n} \log |{\cal L}_n| &=& R_1, 
	\label{eq:rate-R1} \\
\liminf_{n \to \infty} \frac{1}{n} \log |{\cal S}_n| &\ge& R_s.
\end{eqnarray}
Then the achievable region ${\cal R}$ is defined 
as the set of all achievable rate quadruples.
\end{definition}

The following is our main result in this paper.
\begin{theorem}
\label{theorem:main}
Let ${\cal R}^*$ be a closed convex set consisting of 
those quadruples $(R_d,R_0,R_1,R_s)$ for which there
exist auxiliary random variables $(U,V)$ such that 
\begin{eqnarray*}
&& (U,X_2) \markov V \markov X_1, \\
&& (U,V) \markov (X_1,X_2) \markov (Y,Z)
\end{eqnarray*}
and
\begin{eqnarray}
R_0 &\le& \min[I(U,X_2; Y), I(U,X_2; Z)], \label{eq:main-1} \\
R_1 + R_s &\le& I(U,V;Y|X_2), \label{eq:main-2} \\
R_0 + R_1 + R_s &\le& I(V; Y|U,X_2) \nonumber \\
	&& \hspace{-1mm} + \min[I(U,X_2;Y),I(U,X_2;Z)], \label{eq:main-3} \\
 R_s &\le& I(V; Y|U,X_2) - I(V;Z|U,X_2), \label{eq:main-4} \\
 R_1 + R_d &\ge& I(X_1; Z|U,X_2), \label{eq:main-5} \\
 R_d &\ge& I(X_1; Z|U,V,X_2). \label{eq:main-6}
\end{eqnarray}
Then we have ${\cal R} = {\cal R}^*$.
Moreover, it may be assumed that the ranges of $U$
and $V$ may be assumed to satisfy
\begin{eqnarray*}
|{\cal U}| &\le& |{\cal X}_1| |{\cal X}_2| + 3, \\
|{\cal V}| &\le& |{\cal X}_1|^2 |{\cal X}_2|^2 + 4 |{\cal X}_1| |{\cal X}_2| + 3.
\end{eqnarray*}
\end{theorem}
\begin{proof}
See Section \ref{section:proof-main}.
\end{proof}

\begin{remark}
As we will find in the achievability proof of the main theorem,
the private message can be used as dummy randomness to protect
the confidential message from Eve.
Thus, if we define the achievability rate region $\hat{{\cal R}}$ by
replacing Eq.~(\ref{eq:rate-R1}) with
\begin{eqnarray*}
\liminf_{n \to \infty} \frac{1}{n} \log |{\cal L}_n| \ge R_1,
\end{eqnarray*}
region $\hat{{\cal R}}$ is broader than region ${\cal R}$.
Indeed, $\hat{{\cal R}}$ is a closed convex
set consisting of those quadruple $(R_d, R_0, R_1, R_s)$
for which there exist auxiliary random variables $(U,V)$ satisfying
the same conditions as Theorem \ref{theorem:main} except Eq.~(\ref{eq:main-5}).
\end{remark}

\begin{remark}
Eq.~(\ref{eq:main-6}) means that here is a certain amount of dummy randomness
that cannot be substituted by the private message. Note that the difference between
the private message and the dummy randomness is whether Bob needs to 
decode it or not.
\end{remark}

When there is no randomness constraint, region
\begin{eqnarray*}
{\cal R}_{\infty} = \{(R_0,R_1,R_s) : \exists R_d \ge 0~\mbox{s.t.}~(R_d,R_0,R_1,R_s) \in {\cal R} \}
\end{eqnarray*}
coincide with the result obtained by Liang {\em et.~al.} \cite{liang:09}. 
\begin{corollary}
(\cite{liang:09})
Region ${\cal R}_{\infty}$ is a closed convex set consisting of those triplet $(R_0,R_1,R_s)$
for which there exist auxiliary random variables $(U,V)$ such that 
\begin{eqnarray*}
&& (U,X_2) \markov V \markov X_1, \\
&& (U,V) \markov (X_1,X_2) \markov (Y,Z)
\end{eqnarray*}
and
\begin{eqnarray*}
R_0 &\le& \min[I(U,X_2; Y), I(U,X_2; Z)], \\
R_1 + R_s &\le& I(U,V;Y|X_2), \\
R_0 + R_1 + R_s &\le& I(V; Y|U,X_2)  \\
	&& \hspace{-1mm} + \min[I(U,X_2;Y),I(U,X_2;Z)],  \\
 R_s &\le& I(V; Y|U,X_2) - I(V;Z|U,X_2).
\end{eqnarray*}
\end{corollary}

\section{Proof of Main Results}
\label{section:proof-main}

\subsection{Proof of Direct Part of Theorem \ref{theorem:main}}

The direct part of Theorem \ref{theorem:main} follows from
the following Corollary \ref{corollary:direct} and Lemma \ref{lemma-fourier-motzkin}.

We first show the following.
\begin{lemma}
\label{lemma:direct}
Let ${\cal R}^{(in)}$ be a closed convex set consisting of
those quadruples $(R_d,R_0,R_1,R_s)$ for which there exist
$r_1 \ge 0$ and auxiliary random variables $(U,V)$ such that
\begin{eqnarray*}
&& (U,X_2) \markov V \markov X_1, \\
&& (U,V) \markov (X_1,X_2) \markov (Y,Z)
\end{eqnarray*}
and
\begin{eqnarray*}
R_0 + r_1 &\le& I(U,X_2; Z), \\
R_1 - r_1 + R_s &\le& I(V;Y|U,X_2), \\
R_1 + R_s &\le& I(U,V; Y|X_2), \\
R_0 + R_1 + R_s &\le& I(U,V,X_2; Y), \\
R_1 - r_1 &\ge& I(V;Z|U,X_2), \\
R_d &\ge& I(X_1; Z|U,V,X_2).
\end{eqnarray*}
Then we have ${\cal R}^{(in)} \subset {\cal R}$.
\end{lemma}
\begin{proof}
See Section \ref{proof-lemma-direct}.
\end{proof}

We note the following observation.
From the definition of the problem, if
\begin{eqnarray*}
(R_d - r_d, R_0, R_1 - r_s + r_d, R_s + r_s) \in {\cal R}
\end{eqnarray*}
for some $r_d,r_s \ge 0$, then we also have
$(R_d,R_0,R_1,R_s) \in {\cal R}$. Thus, Lemma \ref{lemma:direct}
implies the following corollary.
\begin{corollary}
\label{corollary:direct}
Let $\tilde{{\cal R}}^{(in)}$ be a closed convex set consisting of
those quadruples $(R_d,R_0,R_1,R_s)$ for which there exist
$r_1,r_d,r_s \ge 0$ and $(U,V)$ such that
\begin{eqnarray*}
&& (U,X_2) \markov V \markov X_1, \\
&& (U,V) \markov (X_1,X_2) \markov (Y,Z)
\end{eqnarray*}
and
\begin{eqnarray*}
R_0 + r_1 &\le& I(U,X_2; Z), \\
R_1 - r_1 + r_d + R_s &\le& I(V;Y|U,X_2), \\
R_1 + r_d + R_s &\le& I(U,V; Y|X_2), \\
R_0 + R_1 + r_d + R_s &\le& I(U,V,X_2; Y), \\
R_1 - r_1 - r_s + r_d &\ge& I(V;Z|U,X_2), \\
R_d - r_d &\ge& I(X_1; Z|U,V,X_2).
\end{eqnarray*}
Then we have $\tilde{{\cal R}}^{(in)} \subset {\cal R}$.
\end{corollary}

By using the Fourier-Motzkin elimination, we can also show the following.
\begin{lemma}
\label{lemma-fourier-motzkin}
We have
\begin{eqnarray*}
{\cal R}^* \subset \tilde{{\cal R}}^{(in)}.
\end{eqnarray*}
\end{lemma}
\begin{proof}
See Appendix \ref{proof-lemma-fourier-motzkin}.
\end{proof}

\subsection{Proof of Lemma \ref{lemma:direct}}
\label{proof-lemma-direct}

For a while, we consider the case with $n=1$ and omit the superscript and subscript
to simplify the notation. We first split the private message as
${\cal L} = {\cal I} \times {\cal J}$.
For each common message $k \in {\cal K}$, we randomly 
generate codeword $x_{2k}$ according to distribution $P_{X_2}$.
We denote such a code ${\cal C}_0$. For each $k$ and 
each $i \in {\cal I}$, we randomly generate codeword $u_{ki}$
according to distribution $P_{U|X_2}(\cdot|x_{2k})$.
We denote such a code ${\cal C}_1$.
For each $(k,i)$ and for each $(j,s) \in {\cal J} \times {\cal S}$, we randomly generate 
codeword $v_{kijs}$ according to distribution $P_{V|UX_2}(\cdot|u_{ki},x_{2k})$.
We denote such a code ${\cal C}_2$. For each $(k,i,j)$ and for each 
$a \in {\cal A}$, we randomly generate codeword $x_{1kijsa}$ 
according to distribution $P_{X_1|V}(\cdot|v_{kijs})$. We denote such a code ${\cal C}_3$.

Let 
\begin{eqnarray*}
{\cal T}_0 &=& \left\{ (u,x_2,z) : \frac{P_{Z|UX_2}(z|u,x_2)}{P_Z(z)} \ge e^{\alpha_0} \right\}, \\
{\cal T}_1 &=& \left\{ (u,v,x_2,y) : \frac{P_{Y|UVX_2}(y|u,v,x_2)}{P_{Y|UX_2}(y|u,x_2)} \ge e^{\alpha_1}  \right\}, \\
{\cal T}_2 &=& \left\{ (u,v,x_2,y) : \frac{P_{Y|UVX_2}(y|u,v,x_2)}{P_{Y|X_2}(y|x_2)} \ge e^{\alpha_2}  \right\}, \\
{\cal T}_3 &=& \left\{ (u,v,x_2,y) : \frac{P_{Y|UVX_2}(y|u,v,x_2)}{P_Y(y) } \ge e^{\alpha_3} \right\},
\end{eqnarray*}
and let ${\cal T} = {\cal T}_1 \cap {\cal T}_2 \cap {\cal T}_3$.
Eve decodes only $k$ by using
 the indirect decoding proposed in \cite{nair:09}.
Eve's decoding region is defined by
\begin{eqnarray*}
\lefteqn{{\cal D}_k =} \\
&& \hspace{-3mm} \left\{ z : \exists i~(u_{ik},x_{2i},z) \in {\cal T}_0, \forall \hat{k} \neq k~\forall \hat{i}~(u_{\hat{k}\hat{i}}, x_{2\hat{k}},z) \notin {\cal T}_0 \right\},
\end{eqnarray*}
i.e., $\phi(z) = k$ if $z \in {\cal D}_k$.
Bob decodes $(k,i,j,s)$. Bob's decoding region is defined by
\begin{eqnarray*}
\lefteqn{{\cal D}_{kijs} = \left\{ y : (u_{ki},v_{kijs}, x_{2k}, y) \in {\cal T}, \phantom{\hat{k}v_{\hat{k}\hat{i}\hat{j}\hat{s}}} \right. } \\
&& \left. \forall (\hat{k},\hat{i},\hat{j}, \hat{s} ) \neq (k,i,j,s)~ 
	(u_{\hat{k}\hat{i}}, v_{\hat{k}\hat{i}\hat{j}\hat{s}},x_{2\hat{k}},y) \notin {\cal T} \right\},
\end{eqnarray*}
i.e., $g(y) = (k,i,j,s)$ if $y \in {\cal D}_{kijs}$.

Then we have the following.
\begin{lemma}
\label{lemma:non-asymptotic}
We have
\begin{eqnarray}
\lefteqn{ \mathbb{E}_{{\cal C}_0 {\cal C}_1 {\cal C}_2 {\cal C}_3} \left[ P_{err}(f,g) \right]} \nonumber \\
&\le& P_{UVX_2Y}({\cal T}_1^c) + P_{UVX_2Y}({\cal T}_2^c) + P_{UVX_2Y}({\cal T}_3^c) \nonumber \\
&&	+ |{\cal J}| |{\cal S}| e^{- \alpha_1} + |{\cal I}||{\cal J}||{\cal S}| e^{-\alpha_2} 
	+ |{\cal K}| |{\cal I}| |{\cal J}| |{\cal S}| e^{-\alpha_3},  \nonumber \\
\label{eq:non-asymptotic-bound-1} \\
\lefteqn{ \mathbb{E}_{{\cal C}_0 {\cal C}_1 {\cal C}_2 {\cal C}_3 }\left[ P_{err}(f,\phi) \right] } \nonumber \\
&\le& P_{UX_2Z}({\cal T}_0^c) + |{\cal K}| |{\cal I}| e^{-\alpha_0},
\label{eq:non-asymptotic-bound-2}
\end{eqnarray}
and
\begin{eqnarray}
\lefteqn{ \mathbb{E}_{{\cal C}_0 {\cal C}_1 {\cal C}_2 {\cal C}_3} \left[ D(f) \right]} \nonumber \\
&\le& \frac{1}{ \theta |{\cal A}|^\theta} e^{\psi(\theta|P_{Z|X_1 X_2}, P_{X_1|V},P_{UVX_2})} \nonumber \\
&&	+ \frac{1}{\theta^\prime |{\cal J}|^{\theta^\prime}} e^{\psi(\theta^\prime|P_{Z|UVX_2}, P_{V|UX_2},P_{UX_2})},
\label{eq:non-asymptotic-bound-3}
\end{eqnarray}
where 
\begin{eqnarray*}
\lefteqn{\psi(\theta|P_{Z|X_1 X_2}, P_{X_1|V},P_{UVX_2} )} \\
	&=& \log \sum_{u,v,x_2} P_{UVX_2}(u,v,x_2) \sum_z \\
	&&\left( \sum_{x_1} P_{X_1|V}(x_1|v) P_{Z|X_1 X_2}(z|x_1,x_2)^{1+\theta} \right) \\
	&&	P_{Z|UVX_2}(z|u,v,x_2)^{-\theta}
\end{eqnarray*}
and
\begin{eqnarray*}
\lefteqn{ \psi(\theta^\prime|P_{Z|UVX_2}, P_{V|UX_2},P_{UX_2}) } \\
	&=& \log \sum_{u,x_2} P_{UX_2}(u,x_2) \sum_z \\
	&& \left( \sum_{v} P_{V|U X_2}(v|u,x_2) P_{Z|UVX_2}(z|u,v,x_2)^{1+\theta} \right) \\
	&&	P_{Z|UX_2}(z|u,x_2)^{-\theta}.
\end{eqnarray*}
\end{lemma}
\begin{proof}
See Appendix \ref{proof-of-lemma:non-asymptotic}.
\end{proof}

We apply Lemma \ref{lemma:non-asymptotic} for asymptotic case.
For $(R_d,R_0,R_1 R_s) \in {\cal R}^{(in)}$ 
and arbitrary small $\delta > 0$, we set
$|{\cal K}_n| = \lfloor e^{n(R_0 - \delta)} \rfloor$,
$|{\cal I}_n| = \lfloor e^{n(r_1 - \delta)} \rfloor$,
$|{\cal J}_n| = \lfloor e^{n(R_1 - r_1 + 2 \delta)} \rfloor$,
$|{\cal S}_n| = \lfloor e^{n(R_s - 4 \delta)} \rfloor$,
$|{\cal A}_n| = \lfloor e^{n(R_d + 2 \delta)} \rfloor$,
$\alpha_0 = I(U,X_2;Z) - \delta$, $\alpha_1 = I(V;Y|U,X_2) - \delta$,
$\alpha_2 = I(U,V;Y|X_2) - \delta$, $\alpha_3 = I(U,V,X_2; Y) - \delta$.
Then, 
\begin{eqnarray*}
|{\cal J}_n| |{\cal S}_n| e^{- \alpha_1 n} 
	&\le& e^{-n(I(V;Y|U,X_2) - R_1 + r_1 - R_s + \delta)}, \\
|{\cal I}_n| |{\cal J}_n| |{\cal S}_n| e^{- \alpha_2 n} 
	&\le& e^{-n(I(U,V;Y|X_2) - R_1 - R_s + 2 \delta)}, \\
|{\cal K}_n| |{\cal I}_n| |{\cal J}_n| |{\cal S}_n| e^{-\alpha_3 n} 
	&\le& e^{-n(I(U,V,X_2; Y) - R_0 - R_1 - R_s + 3 \delta)}, \\
|{\cal K}_n| |{\cal I}_n| e^{- \alpha_0 n}
	&\le& e^{-n(I(U,X_2; Z) - R_0 -r_1 + \delta)}
\end{eqnarray*}
converge to $0$ asymptotically.
Furthermore, by the law of large numbers, 
$P_{UVX_2Y}^n({\cal T}_{1,n}^c)$,
$P_{UVX_2Y}^n({\cal T}_{2,n}^c)$,
$P_{UVX_2Y}^n({\cal T}_{3,n}^c)$,
and $P_{UX_2Z}^n({\cal T}_{0,n}^c)$ also converge to $0$ asymptotically.

Since 
\begin{eqnarray*}
\psi^\prime(0|P_{Z|X_1 X_2},P_{X_1|V},P_{UVX_2}) = I(X_1;Z|U,V,X_2)
\end{eqnarray*}
there exists $\theta_0 > 0$ such that
\begin{eqnarray*}
\lefteqn{\frac{\psi(\theta_0| P_{Z|X_1 X_2},P_{X_1|V},P_{UVX_2})}{\theta_0} } \\
&\le& I(X_1;Z|U,V,X_2) + \delta \le R_d + \delta,
\end{eqnarray*}
which implies 
\begin{eqnarray*}
- \frac{\theta_0}{n} \log |{\cal A}_n| + \psi(\theta_0| P_{Z|X_1 X_2},P_{X_1|V},P_{UVX_2}) \le \delta.
\end{eqnarray*}
Thus, 
\begin{eqnarray*}
\frac{1}{\theta_0 |{\cal A}_n|^{\theta_0}} e^{n\psi(\theta_0| P_{Z|X_1 X_2},P_{X_1|V},P_{UVX_2})}
\end{eqnarray*}
exponentially converges to $0$.
Similarly, since 
\begin{eqnarray*}
\psi^\prime(0|P_{Z|UVX_2},P_{V|UX_2},P_{UX_2}) = I(V;Z|U,X_2)
\end{eqnarray*}
there
exists $\theta_0^\prime > 0$ such that
\begin{eqnarray*}
\lefteqn{\frac{\psi(\theta_0^\prime| P_{Z|UVX_2},P_{V|UX_2},P_{UX_2})}{\theta_0^\prime} } \\
	&\le& I(V;Z|U,X_2) + \delta \le R_1 - r_1 + \delta,
\end{eqnarray*}
which implies 
\begin{eqnarray*}
- \frac{\theta_0^\prime}{n} \log |{\cal J}_n| + \psi(\theta_0^\prime| P_{Z|UVX_2},P_{V|UX_2},P_{UX_2}) \le - \delta.
\end{eqnarray*}
Thus, 
\begin{eqnarray*}
\frac{1}{\theta_0^\prime |{\cal J}_n|^{\theta_0^\prime}} e^{n \psi(\theta_0^\prime| P_{Z|UVX_2},P_{V|UX_2},P_{UX_2})}
\end{eqnarray*}
exponentially converges to $0$ asymptotically.
This completes a proof of the lemma. \qed

\subsection{Proof of Converse Part of Theorem \ref{theorem:main}}

Suppose that $(R_d, R_0, R_1, R_s) \in {\cal R}$. Then, for arbitrary $\gamma >0$, 
there exists $n$ such that 
\begin{eqnarray*}
n(R_0 - \gamma) &\le& \log |{\cal K}_n|, \\
n(R_1 + R_s - \gamma) &\le& \log |{\cal L}_n||{\cal S}_n|, \\
n(R_0 + R_1 + R_s - \gamma) &\le& \log |{\cal K}_n||{\cal L}_n||{\cal S}_n|, \\
n(R_s - \gamma) &\le& \log |{\cal S}_n|, \\
n(R_1 + R_d + \gamma) &\ge& \log |{\cal L}_n| |{\cal A}_n|, \\
n(R_d - \gamma) &\ge& \log |{\cal A}_n|.
\end{eqnarray*}
By combining these inequalities with the following
Lemma \ref{lemma:fano} and Lemma \ref{lemma:single-letter}, we have the 
converse part of the theorem.
The statement about the range size of $U$ and $V$ can be
proved in the same manner as \cite{liang:09}.
It should be noted that Eqs.~(\ref{eq:main-1})--(\ref{eq:main-4}) are 
derived in the same manner as \cite{liang:09} and the 
construction  of the auxiliary random variable
are also the same. Eqs.~(\ref{eq:main-5}) and (\ref{eq:main-6}) are
additionally proved in this paper by using the fact that
Alice's encoder is deterministic given the dummy randomness. 

\begin{lemma}
\label{lemma:fano}
There exists $\varepsilon_n \to 0$ such that
\begin{eqnarray*}
\lefteqn{\log |{\cal K}_n| } \\
	&\le& I(K_n, X_2^n; Y^n) + n \varepsilon_n, \\
\lefteqn{ \log |{\cal K}_n| } \\
	&\le& I(K_n,X_2^n; Z^n) + n \varepsilon_n, \\
\lefteqn{ \log |{\cal L}_n| |{\cal S}_n| } \\
	&\le& I(L_n, S_n; Y^n | K_n, X_2^n) + n \varepsilon_n, \\
\lefteqn{ \log |{\cal K}_n| |{\cal L}_n| |{\cal S}_n| } \\
	&\le& I(K_n,L_n,S_n,X_2^n;Y^n) + n \varepsilon_n, \\
\lefteqn{ \log |{\cal K}_n| |{\cal L}_n| |{\cal S}_n| } \\
	&\le& I(L_n,S_n; Y^n | K_n,X_2^n) + I(K_n,X_2^n; Z^n) + 2 n \varepsilon_n, \\
\lefteqn{ \log |{\cal S}_n| } \\
	&\le& I(L_n,S_n; Y^n | K_n, X_2^n)  \\
	&& - I(L_n, S_n; Z^n | K_n, X_2^n) + 4 n \varepsilon_n, \\
\lefteqn{ \log |{\cal L}_n| |{\cal A}_n| } \\
	&\ge& I(X_1^n; Z^n | K_n, X_2^n) - 2 n \varepsilon_n, \\
\lefteqn{ \log |{\cal A}_n| } \\
	&\ge& I(X_1^n ; Z^n | K_n, L_n, S_n, X_2^n).
\end{eqnarray*}
\end{lemma}
\begin{proof}
By using Fano's inequality, we have
\begin{eqnarray*}
\log |{\cal K}_n| 
	&=& H(K_n) \\
	&=& I(K_n; Y^n) + H(K_n|Y^n) \\
	&\le& I(K_n, X_2^n; Y^n) + n \varepsilon_n, \\
\end{eqnarray*}
and
\begin{eqnarray*}
\log |{\cal K}_n| \le I(K_n,X_2^n; Z^n) + n \varepsilon_n.
\end{eqnarray*}
By using Fano's inequality and by noting that 
$(K_n,X_2^n)$ and $(L_n,S_n)$ are independent, we have
\begin{eqnarray*}
\log |{\cal L}_n| |{\cal S}_n|
	&=& H(L_n,S_n) \\
	&=& I(L_n,S_n; Y^n) + H(L_n,S_n|Y^n) \\
	&\le& I(L_n, S_n; K_n, X_2^n,Y^n) + n \varepsilon_n \\
	&=& I(L_n, S_n; Y^n|K_n,X_2^n) + n \varepsilon_n.
\end{eqnarray*}
By using Fano's inequality, we also have
\begin{eqnarray*}
\log |{\cal K}_n| |{\cal L}_n| |{\cal S}_n| 
	&=& H(K_n,L_n,S_n) \\
	&\le& I(K_n,L_n,S_n,X_2^n; Y^n) + n \varepsilon_n
\end{eqnarray*}
and
\begin{eqnarray*}
\lefteqn{ \log |{\cal K}_n| |{\cal L}_n| |{\cal S}_n|  } \\
	&=& H(L_n, S_n|K_n) + H(K_n) \\
	&\le& I(L_n, S_n; Y^n| K_n) + I(K_n,X_2^n; Z^n) + 2 n \varepsilon_n \\
	&=& I(L_n,S_n; Y^n |K_n,X_2^n) + I(K_n,X_2^n; Z^n) + 2 n \varepsilon_n,
\end{eqnarray*}
where the last equality follows from the fact that $X_2^n$ is a determined from $K_n$.
By using the security condition and Fano's inequality, we have
\begin{eqnarray}
\lefteqn{ I(S_n; Z^n|K_n) } \nonumber \\
	&=& I(S_n,K_n ; Z^n) - I(K_n;Z^n) \nonumber \\
	&=& I(S_n; Z^n) + I(K_n;Z^n|S_n) - I(K_n;Z^n) \nonumber \\
	&\le& I(S_n;Z^n) + H(K_n|Z^n) \nonumber \\
	&\le& 2 n \varepsilon_n.
	\label{eq:szk-bound}
\end{eqnarray}
By using Fano's inequality and
by using Eq.~(\ref{eq:szk-bound}), we have
\begin{eqnarray*}
\lefteqn{ \log |{\cal S}_n| } \\
	&=& H(S_n|K_n) \\
	&\le& I(S_n; Y^n|K_n) + n \varepsilon_n \\
	&=& I(L_n,S_n; Y^n|K_n) - I(L_n;Y^n|S_n,K_n) + n \varepsilon_n \\
	&\le& I(L_n,S_n; Y^n|K_n) - H(L_n|S_n,K_n) + 2 n \varepsilon_n \\
	&\le& I(L_n,S_n; Y^n|K_n) - I(S_n;Z^n|K_n) \\
	&& - H(L_n|S_n,K_n) + 4 n \varepsilon_n \\ 
	&\le& I(L_n,S_n; Y^n|K_n) - I(L_n,S_n;Z^n|K_n) + 4 n \varepsilon_n \\
	&=& I(L_n,S_n; Y^n|K_n,X_2^n) \\
	&& - I(L_n,S_n;Z^n|K_n,X_2^n) + 4 n \varepsilon_n.
\end{eqnarray*}
By noting that $f_n$ is a deterministic function
and by using Eq.~(\ref{eq:szk-bound}), we have
\begin{eqnarray*}
\lefteqn{ \log |{\cal L}_n| |{\cal A}_n| } \\
	&\ge& H(X_1^n|K_n,S_n) \\
	&\ge& I(X_1^n; Z^n| K_n, S_n) \\
	&=& I(X_1^n, S_n ; Z^n|K_n) - I(S_n; Z^n|K_n) \\
	&\ge& I(X_1^n; Z^n|K_n) - 2 n \varepsilon_n \\
	&=& I(X_1^n; Z^n|K_n,X_2^n) - 2 n \varepsilon_n.
\end{eqnarray*}
Finally, by noting that $f_n$ is a deterministic function, we have
\begin{eqnarray*}
\log |{\cal A}_n| 
	&\ge& H(X_1^n|K_n,L_n,S_n) \\
	&\ge& I(X_1^n; Z^n|K_n, L_n, S_n) \\
	&=& I(X_1^n; Z^n|K_n, L_n, S_n,X_2^n).
\end{eqnarray*}
\end{proof}

\begin{lemma}
\label{lemma:single-letter}
For fixed $n$, let $T$ be the random variable that is uniformly 
distributed on $\{1,\ldots,n\}$ and is independent of the other 
random variables. Define the following random variables:
\begin{eqnarray*}
U_t &=& (K_n,X_2^n,Y_1^{t-1},Z_{t+1}^n), \\
V_t &=& (L_n,S_n,U_t), \\
U &=& (U_T,T), \\
V &=& (V_T,T), \\
X_1 &=& X_{1T}, \\
X_2 &=& X_{2T}, \\
Y &=& Y_T, \\
Z &=& Z_T.
\end{eqnarray*}
Then, we have
\begin{eqnarray}
\lefteqn{ I(K_n,X_2^n; Y^n) } \nonumber \\
	&\le& n I(U, X_2; Y), \label{eq:single-letter-1} \\
\lefteqn{ I(K_n, X_2^n; Z^n) } \nonumber \\
	&\le& n I(U,X_2; Z), \label{eq:single-letter-2} \\
\lefteqn{ I(L_n, S_n; Y^n | K_n, X_2^n) } \nonumber \\
	&\le& n I(U,V; Y| X_2), \label{eq:single-letter-3} \\
\lefteqn{ I(K_n, L_n,S_n,X_2^n ; Y^n) } \nonumber \\
	&\le& n I(U,V,X_2; Y), \label{eq:single-letter-4} \\
\lefteqn{ I(L_n,S_n; Y^n | K_n, X_2^n) + I(K_n, X_2^n ; Z^n) } \nonumber \\
	&\le& n[ I(V; Y|U, X_2) + I(U, X_2; Z)], \label{eq:single-letter-5} \\
\lefteqn{ I(L_n, S_n; Y^n | K_n, X_2^n) - I(L_n, S_n; Z^n | K_n, X_2^n) } \nonumber \\
	&\le& n[ I(V; Y|U, X_2) - I(V; Z|U, X_2)], \label{eq:single-letter-6} \\
\lefteqn{ I(X_1^n; Z^n | K_n, X_2^n) } \nonumber \\
	&\ge& n I(X_1; Z|U, X_2), \label{eq:single-letter-7} \\
\lefteqn{ I(X_1^n; Z^n | K_n, L_n, S_n, X_2^n) } \nonumber \\
	&\ge& n I(X_1; Z|U, V, X_2). \label{eq:single-letter-8}
\end{eqnarray}
\end{lemma}
\begin{proof}
\paragraph{Proof of Eq.~(\ref{eq:single-letter-1})}
\begin{eqnarray*}
\lefteqn{ I(K_n,X_2^n; Y^n) } \\
	&=& \sum_{t=1}^n I(K_n,X_2^n; Y_t| Y_1^{t-1}) \\
	&\le& \sum_{t=1}^n I(K_n,X_2^n,Y_1^{t-1},Z_{t+1}^n; Y_t) \\
	&=& \sum_{t=1}^n I(U_t; Y_t) \\
	&=& n I(U_T; Y_T|T) \\
	&=& n I(U_T,T; Y_T) \\
	&=& n I(U; Y).
\end{eqnarray*}
\paragraph{Proof of Eq.~(\ref{eq:single-letter-2})}
\begin{eqnarray*}
\lefteqn{ I(K_n,X_2^n; Z^n) } \\
	&=& \sum_{t=1}^n I(K_n,X_2^n; Z_t| Z_{t+1}^n) \\
	&\le& \sum_{t=1}^n I(K_n,X_2^n,Y_1^{t-1},Z_{t+1}^n; Z_t) \\
	&=& \sum_{t=1}^n I(U_t; Z_t) \\
	&=& n I(U_T; Z_T|T) \\
	&=& n I(U_T,T; Z_T) \\
	&=& n I(U; Z).
\end{eqnarray*}
\paragraph{Proof of Eq.~(\ref{eq:single-letter-3})}
\begin{eqnarray*}
\lefteqn{ I(L_n,S_n ; Y^n | K_n,X_2^n) } \\
	&=& \sum_{t=1}^n [ H(Y_t|K_n,X_2^n,Y_1^{t-1}) \\
	&& - H(Y_t|K_n,L_n,S_n,X_2^n,Y_1^{t-1})] \\
	&\le& \sum_{t=1}^n[ H(Y_t|X_{2t}) \\
	&& - H(Y_t|K_n,L_n,S_n,X_2^n,Y_1^{t-1},Z_{t+1}^n) ] \\
	&=& \sum_{t=1}^n I(K_n,L_n,S_n,X_2^n; Y_1^{t-1},Z_{t+1}^n ; Y_t | X_{2t}) \\
	&=& \sum_{t=1}^n I(U_t, V_t ; Y_t | X_{2t}) \\
	&=& n I(U_T,V_T; Y_T | X_{2T},T) \\
	&=& n I(U, V; Y| X_2).
\end{eqnarray*}
\paragraph{Proof of Eq.~(\ref{eq:single-letter-4})}
\begin{eqnarray*}
\lefteqn{ I(K_n,L_n,S_n,X_2^n ; Y^n) } \\
	&=& \sum_{t=1}^n I(K_n, L_n, S_n, X_2^n ; Y_t | Y_1^{t-1}) \\
	&\le& \sum_{t=1}^n I(K_n, L_n, S_n, X_2^n, Y_1^{t-1}, Z_{t+1}^n ; Y_t) \\
	&=& \sum_{t=1}^n I(U_t,V_t,X_{2t} ; Y_t) \\
	&=& n I(U_T,V_T,X_{2T}; Y_T | T) \\
	&=& n I(U,V,X_2; Y).
\end{eqnarray*}
\paragraph{Proof of Eq.~(\ref{eq:single-letter-5})}
\begin{eqnarray*}
\lefteqn{ I(L_n,S_n ; Y^n|K_n,X_2^n) + I(K_n,X_2^n ; Z^n) } \\
	&=& \sum_{t=1}^n [ I(L_n,S_n; Y_t | K_n, X_2^n, Y_1^{t-1}) \\
	&& + I(K_n, X_2^n ; Z_t | Z_{t+1}^n ) ] \\
	&\le& \sum_{t=1}^n [ I(L_n,S_n,Z_{t+1}^n ; Y_t | K_n, X_2^n,Y_1^{t-1}) \\
	&& - I(Y_1^{t-1} ; Z_t | K_n, X_2^n, Z_{t+1}^n) \\
	&& + I(K_n, X_2^n, Y_1^{t-1} ; Z_t | Z_{t+1}^n)] \\
	&=& \sum_{t=1}^n [ I(L_n, S_n ; Y_t | K_n, X_2^n, Y_1^{t-1}, Z_{t+1}^n) \\
	&& + I(Z_{t+1}^n ; Y_t | K_n, X_2^n, Y_1^{t-1}) \\
	&& - I(Y_1^{t-1} ; Z_t | K_n, X_2^n, Z_{t+1}^n) \\
	&& + I(K_n,X_2^n,Y_1^{t-1} ; Z_t | Z_{t+1}^n)] \\
	&\stackrel{(a)}{=}& \sum_{t=1}^n[ I(L_n,S_n ; Y_t | K_n,X_2^n,Y_1^{t-1},Z_{t+1}^n) \\
	&& + I(K_n,X_2^n, Y_1^{t-1} ; Z_t | Z_{t+1}^n) ] \\
	&\le& \sum_{t=1}^n[ I(L_n,S_n ; Y_t | K_n,X_2^n,Y_1^{t-1},Z_{t+1}^n) \\
	&& + I(K_n,X_2^n, Y_1^{t-1}, Z_{t+1}^n ; Z_t) ] \\
	&=& n[I(V_T;Y_T|U_T,X_{2T},T) + I(U_T,X_{2T}; Z_T|T)] \\
	&=& n[I(V;Y|U,X_2) + I(U,X_2;Z)],
\end{eqnarray*}
where we used Csisz\'ar's sum identity \cite{elgamal-kim-book} in (a).
\paragraph{Proof of Eq.~(\ref{eq:single-letter-6})}
\begin{eqnarray*}
\lefteqn{ I(L_n,S_n ; Y^n | K_n,X_2^n) - I(L_n,S_n ; Z^n | K_n,X_2^n) } \\
	&=& \sum_{t=1}^n [ I(L_n,S_n ; Y_t | K_n,X_2^n,Y_1^{t-1}) \\
	&& - I(L_n,S_n ; Z_t | K_n,X_2^n,Z_{t+1}^n)] \\
	&\stackrel{(a)}{=}& \sum_{t=1}^n[ I(L_n,S_n ; Y_t | K_n,X_2^n,Y_1^{t-1}) \\
	&& + I(Z_{t+1}^n; Y_t|K_n,L_n,S_n,X_2^n,Y_1^{t-1}) \\
	&& - I(Y_1^{t-1} ; Z_t | K_n, L_n, S_n, X_2^n,Z_{t+1}^n) \\
	&& - I(L_n,S_n ; Z_t | K_n, X_2^n, Z_{t+1}^n) \\
	&=& \sum_{t=1}^n[ I(L_n,S_n,Z_{t+1}^n ; Y_t | K_n, X_2^n, Y_1^{t-1}) \\
	&& - I(L_n,S_n,Y_1^{t-1} ; Z_t | K_n, X_2^n, Z_{t+1}^n)] \\
	&=& \sum_{t=1}^n[ I(L_n,S_n ; Y_t | K_n, X_2^n, Y_1^{t-1},Z_{t+1}^n) \\
	&& + I(Z_{t+1}^n ; Y_t | K_n,X_2^n,Y_1^{t-1}) \\
	&& - I(Y_1^{t-1}; Z_t | K_n, X_2^n, Z_{t+1}^n) \\
	&& - I(L_n, S_n; Z_t | K_n, X_2^n, Y_1^{t-1},Z_{t+1}^n)] \\
	&\stackrel{(b)}{=}& \sum_{t=1}^n[ I(L_n,S_n ; Y_t | K_n, X_2^n,Y_1^{t-1},Z_{t+1}^n) \\
	&& - I(L_n,S_n ; Z_t | K_n, X_2^n,Y_1^{t-1},Z_{t+1}^n)] \\
	&=& n[I(V_T;Y_T|U_T,X_{2T},T) - I(V_T;Z_T|U_T,X_{2T},T)] \\
	&=& n[I(V;Y|U,X_2) - I(V;Z|U,X_2)],
\end{eqnarray*}
where (a) and (b) follow from Csisz\'ar's sum identity \cite{elgamal-kim-book}.
\paragraph{Proof of Eq.~(\ref{eq:single-letter-7})}
\begin{eqnarray*}
\lefteqn{ I(X_1^n ; Z^n | K_n,X_2^n) } \\
	&=& \sum_{t=1}^n[ H(Z_t|K_n,X_2^n,Z_{t+1}^n)  \\
	&& - H(Z_t|K_n,X_1^n,X_2^n,Z_{t+1}^n)] \\
	&\stackrel{(a)}{\ge}& \sum_{t=1}^n[ H(Z_t|K_n,X_2^n,Y_1^{t-1},Z_{t+1}^n) \\
	&& - H(Z_t|K_n,X_{1t},X_2^n,Y_1^{t-1},Z_{t+1}^n)] \\
	&=& \sum_{t=1}^n I(X_{1t} ; Z_t | K_n,X_2^n,Y_1^{t-1},Z_{t+1}^n) \\
	&=& \sum_{t=1}^n I(X_{1t} ; Z_t | U_t,X_{2t}) \\
	&=& n I(X_{1T} ; Z_T | U_T,X_{2T},T) \\
	&=& n I(X_1 ; Z |U,X_2),
\end{eqnarray*}
where (a) follows from the fact that $(K_n$, $X_{11}^{t-1}$, $X_{1(t+1)}^n$, $X_{21}^{t-1}$, $X_{2(t+1)}^n$, $Y_1^{t-1}$, $Z_{t+1})$, $(X_{1t}$,$X_{2t})$, and 
$Z_t$ form Markov chain.
\paragraph{Proof of Eq.~(\ref{eq:single-letter-8})}
\begin{eqnarray*}
\lefteqn{ I(X_1^n; Z^n | K_n,L_n,S_n,X_2^n)  } \\
	&=& \sum_{t=1}^n[ H(Z_t|K_n, L_n, S_n,X_2^n,Z_{t+1}^n) \\
	&& - H(Z_t|K_n,L_n,S_n,X_1^n,X_2^n,Z_{t+1}^n) ] \\
	&\stackrel{(a)}{\ge}& \sum_{t=1}^n[ H(Z_t|K_n,L_n,S_n,X_2^n,Y_1^{t-1},Z_{t+1}^n) \\
	&& - H(Z_t|K_n,L_n,S_n,X_{1t},X_2^n,Y_1^{t-1},Z_{t+1}^n)] \\
	&=& \sum_{t=1}^n I(X_{1t}; Z_t |K_n,L_n,S_n,X_2^n,Y_1^{t-1},Z_{t+1}^n) \\
	&=& \sum_{t=1}^n I(X_{1t}; Z_t | U_t,V_t,X_{2t}) \\
	&=& n I(X_{1T} ; Z_T | U_T,V_T,X_{2T},T) \\
	&=& n I(X_1; Z | U, V, X_2),
\end{eqnarray*}
where (a) follows from the fact that $(K_n$, $L_n$, $S_n$, $X_{11}^{t-1}$, $X_{1(t+1)}^n$, $X_{21}^{t-1}$, $X_{2(t+1)}^n$, $Y_1^{t-1}$, $Z_{t+1})$, $(X_{1t}$,$X_{2t})$, and 
$Z_t$ form Markov chain.
\end{proof}


\section*{Acknowledgment}

The authors would like to thank Prof.~Ryutaroh Matsumoto
for teaching them the practical importance of the randomness
constraint. This research is partly supported by Grant-in-Aid
for Young Scientist(B):2376033700,
Grand-in-Aid for Scientific Research(B):2336017202,
and Grand-in-Aid for Scientific Research(A):2324607101.

\appendix

\subsection{Channel Resolvability}

Since we use a result of the channel resolvability problem \cite{han:93}
in the proof of our main result, we review the channel resolvability
problem in this appendix. For simplicity of notation, we consider the
so-called one-shot case, i.e., the block length is $n=1$.
In the channel resolvability problem, for the input distribution
$P_X$ of the channel $P_{Z|X}$, we want to simulate the response
$P_Z$ of the channel, where 
\begin{eqnarray*}
P_Z(z) = \sum_x P_X(x) P_{Z|X}(z|x).
\end{eqnarray*}
The simulation is conducted by a deterministic map
$\varphi:{\cal B} \to {\cal X}$, and uniform random number $B$ on ${\cal B}$.
Let 
\begin{eqnarray*}
P_{\tilde{Z}}(z) = \sum_{b \in {\cal B}} \frac{1}{|{\cal B}|} P_{Z|X}(z|\varphi(b))
\end{eqnarray*}
be the output distribution with map $\varphi$. The purpose of the resolvability
problem is to construct a map such that $D(P_{\tilde{Z}} \| P_Z)$ is small.

In \cite{watanabe:12}, the following random coding construction of a map was proposed.
We split the alphabet as ${\cal B} = {\cal M}_1 \times {\cal M}_2$.
Let $P_{VX}$ be a distribution such that the marginal is $P_X$.
We first randomly generate $|{\cal M}_2|$ codewords 
$v_1,\ldots, v_{|{\cal M}_2|}$ according to the distribution
$P_V$. We denote the generated code by ${\cal C}_2$.
Then, for each $1 \le i \le |{\cal M}_2|$, we randomly generate
$|{\cal M}_1|$ codewords $x_{i1},\ldots,x_{i |{\cal M}_1|}$
according to the distribution $P_{X|V}(\cdot | v_i)$.
We denote the generated code by ${\cal C}_1$.
For this construction we have the following lemma.
\begin{lemma}
(\cite{watanabe:12})
\label{lemma:superposition-resolvability}
For $0 < \theta, \theta^\prime \le 1$, we have
\begin{eqnarray*}
\lefteqn{ \mathbb{E}_{{\cal C}_1 {\cal C}_2} \left[  D(P_{\tilde{Z}} \| P_Z) \right] } \\
&\le& \frac{1}{\theta |{\cal M}_1|^\theta} e^{\psi(\theta | P_{Z|X}, P_{X|V},P_V)} \\
&& + \frac{1}{\theta^\prime |{\cal M}_2|^{\theta^\prime}} e^{\psi( \theta^\prime | P_{Z|V},P_V)},
\end{eqnarray*}
where
\begin{eqnarray*}
\lefteqn{ \psi(\theta | P_{Z|X}, P_{X|V}, P_V) } \\
&=& \log \sum_v P_V(v) \sum_z  \\
&& \left( \sum_x P_{X|V}(x|v) P_{Z|X}(z|x)^{1+\theta} \right) P_{Z|V}(z|v)^{-\theta}
\end{eqnarray*}
and
\begin{eqnarray*}
\lefteqn{ \psi( \theta^\prime | P_{Z|V},P_V)  } \\
&=& \log \sum_z \left( \sum_x P_X(x) P_{Z|X}(z|x)^{1+\theta^\prime} \right) P_Z(x)^{- \theta^\prime}.
\end{eqnarray*}
\end{lemma} 

\subsection{Proof of Lemma \ref{lemma:non-asymptotic}}
\label{proof-of-lemma:non-asymptotic}

\paragraph{Proof of Eq.~(\ref{eq:non-asymptotic-bound-1})}
We first not the following observation.
By taking the average over randomly generated codes, we have
\begin{eqnarray}
\lefteqn{ \mathbb{E}_{{\cal C}_0 {\cal C}_1 {\cal C}_2 {\cal C}_3}\left[ P_{err}(f,g) \right] } \nonumber \\
&=& \mathbb{E}_{{\cal C}_0 {\cal C}_1 {\cal C}_2 {\cal C}_3}\left[ 
	\sum_{k,i,j,s,a} \frac{1}{|{\cal K}| |{\cal I}| |{\cal J}| |{\cal S}| |{\cal A}|} \right. \nonumber \\
&& \left. \phantom{\sum_{k,i,j,s,a} \frac{1}{|{\cal K}|}}	P_{Y|X_1 X_2}({\cal D}_{kijsa}^c|x_{1kijsa}, x_{2k})
 \right] \nonumber \\
&=& \mathbb{E}_{{\cal C}_0 {\cal C}_1 {\cal C}_2}\left[ 
	\sum_{k,i,j,s,a} \frac{1}{|{\cal K}| |{\cal I}| |{\cal J}| |{\cal S}| |{\cal A}|} \right. \nonumber \\
&& \left. \phantom{\sum_{k,i,j,s,a} \frac{1}{|{\cal K}|}}	\mathbb{E}_{{\cal C}_3}\left[ P_{Y|X_1 X_2}({\cal D}_{kijsa}^c|x_{1kijsa}, x_{2k}) \right]
 \right] \nonumber \\
&=& \mathbb{E}_{{\cal C}_0 {\cal C}_1 {\cal C}_2 }\left[ 
	\sum_{k,i,j,s} \frac{1}{|{\cal K}| |{\cal I}| |{\cal J}| |{\cal S}| } \right. \nonumber \\
&& \left. \phantom{\sum_{k,i,j,s,a} \frac{1}{|{\cal K}|}}	P_{Y|V X_2}({\cal D}_{kijsa}^c|v_{kijs}, x_{2k})
 \right].
 \label{eq:proof-non-asymptotic-1}
\end{eqnarray}
Let ${\cal T}_{uvx_2} = \{ y : (u,v,x_2,y) \in {\cal T} \}$.
Then, we have
\begin{eqnarray*}
\lefteqn{  \mathbb{E}_{{\cal C}_0 {\cal C}_1 {\cal C}_2 }\left[ 
	\sum_{k,i,j,s} \frac{1}{|{\cal K}| |{\cal I}| |{\cal J}| |{\cal S}| }
	P_{Y|V X_2}({\cal D}_{kijsa}^c|v_{kijs}, x_{2k})
 \right]. } \\
 &\le& \mathbb{E}_{{\cal C}_0 {\cal C}_1 {\cal C}_2} \left[ 
 	\sum_{k,i,j,s} \frac{1}{ |{\cal K}| |{\cal I}| |{\cal J}| |{\cal S}| }  \right. \\
&&	\{ P_{Y|VX_2}({\cal T}_{u_{ki}v_{kijs} x_{2k}}^c| v_{kijs},x_{2k})  \\
&& \left. + \sum_{(\hat{k},\hat{i},\hat{j},\hat{s}) \atop \neq (k,i,j,s)} 
	P_{Y|VX_2}({\cal T}_{u_{\hat{k}\hat{i}} v_{\hat{k}\hat{i}\hat{j}\hat{s}}x_{2\hat{k}}} | v_{kijs},x_{2k}) \}	\right] \\
&\le& \mathbb{E}_{{\cal C}_0 {\cal C}_1 {\cal C}_3} \left[ \sum_{k,i,j,s} \frac{1}{|{\cal K}| |{\cal I}| |{\cal J}| |{\cal S}|} \right. \\
&&	\{ P_{Y|VX_2} ({\cal T}_{u_{ki}v_{kijs} x_{2k}}^c| v_{kijs},x_{2k})  \\
&& + \sum_{(\hat{j},\hat{s}) \neq (j,s)} P_{Y|VX_2}({\cal T}_{u_{ki} v_{ki\hat{j}\hat{s}} x_{2k}} | v_{kijs}, x_{2k}) \\
&& + \sum_{\hat{i} \neq i} \sum_{(\hat{j},\hat{s})} P_{Y|VX_2}({\cal T}_{u_{k\hat{i}} v_{k\hat{i}\hat{j}\hat{s}} x_{2k}} | v_{kijs}, x_{2k}) \\
&& \left. + \sum_{\hat{k} \neq k} \sum_{(\hat{i},\hat{j},\hat{s})} P_{Y|VX_2}({\cal T}_{u_{\hat{k}\hat{i}} v_{\hat{k}\hat{i}\hat{j}\hat{s}} x_{2\hat{k}}}| v_{kijs},x_{2k}) 
\} 	\right] \\
&\le& \sum_{k,i,j,s} \frac{1}{|{\cal K}| |{\cal I}| |{\cal J}| |{\cal S}| } \{ P_{UVX_2Y}({\cal T}^c) \\
&& + |{\cal J}| |{\cal S}| \sum_{u,v,x_2} P_{UVX_2}(u,v,x_2) P_{Y|UX_2}({\cal T}_{uvx_2} | u,x_2) \\
&& + |{\cal I}| |{\cal J}| |{\cal S}| \sum_{u,v,x_2} P_{UVX_2}(u,v,x_2) P_{Y|X_2}({\cal T}_{uvx_2} | x_2) \\
&& + |{\cal K}| |{\cal I}| |{\cal J}| |{\cal S}| \sum_{u,v,x_2} P_{UVX_2}(u,v,x_2) P_Y({\cal T}_{uvx_2}) \\
&\le& P_{UVX_2Y}({\cal T}^c) + |{\cal J}| |{\cal S}| e^{-\alpha_1} \\
&& + |{\cal I}| |{\cal J}| |{\cal S}| e^{-\alpha_2} + |{\cal K}| |{\cal I}| |{\cal J}| |{\cal S}| e^{-\alpha_3},
\end{eqnarray*}
where we used 
\begin{eqnarray*}
P_{Y|UX_2}(y|u,x_2) &\le& P_{Y|UVX_2}(y|u,v,x_2) e^{-\alpha_1}, \\
P_{Y|X_2}(y|x_2) &\le& P_{Y|UVX_2}(y|u,v,x_2) e^{- \alpha_2}, \\
P_Y(y) &\le& P_{Y|UVX_2}(y|u,v,x_2) e^{- \alpha_3}
\end{eqnarray*}
for $y \in {\cal T}_{uvx_2}$ in the last inequality.

\paragraph{Proof of Eq.~(\ref{eq:non-asymptotic-bound-2})}

Let ${\cal T}_{0,ux_2} = \{ z : (u,x_2,z) \in {\cal T}_0 \}$.
In a similar manner as Eq.~(\ref{eq:proof-non-asymptotic-1}), we have
\begin{eqnarray*}
\lefteqn{ \mathbb{E}_{{\cal C}_0 {\cal C}_1 {\cal C}_2 {\cal C}_3}\left[ P_{err}(f,\phi) \right]} \\
&=& \mathbb{E}_{{\cal C}_0 {\cal C}_1}\left[ \sum_{k,i} \frac{1}{|{\cal K}| |{\cal I}|} 
	P_{Z|UX_2}({\cal D}_k^c | u_{ki},x_{2k}) \right] \\
&\le& \mathbb{E}_{{\cal C}_0 {\cal C}_1} \left[ \sum_{k,i} \frac{1}{|{\cal K}| |{\cal I}|} \{
	P_{Z|UX_2}({\cal T}_{0,u_{ki} x_{2k}}^c | u_{ki},x_{2k}) \right. \\
&& + \left. \sum_{\hat{k} \neq k} \sum_{\hat{i}} P_{Z|UX_2}({\cal T}_{0,u_{\hat{k}\hat{i}} x_{2\hat{k}}} | u_{ki}, x_{2k}) \} \right] \\
&\le& \sum_{k,i} \frac{1}{|{\cal K}| |{\cal I}|} \{ P_{UX_2Z}({\cal T}_0^c) \\
&& + |{\cal K}| |{\cal I}| \sum_{u,x_2} 
	P_{UX_2}(u,x_2) P_Z({\cal T}_{0,u x_2}) \} \\
&\le& P_{UX_2Z}({\cal T}_0^c) + |{\cal K}| |{\cal I}| e^{- \alpha_0},
\end{eqnarray*}
where we used
\begin{eqnarray*}
P_Z(z) \le P_{Z|UX_2}(z|u,x_2) e^{- \alpha_0}
\end{eqnarray*}
for $z \in {\cal T}_{0,u x_2}$ in the last inequality.

\paragraph{Proof of Eq.~(\ref{eq:non-asymptotic-bound-3})}

By using the monotonicity of the divergence, we have
\begin{eqnarray*}
D(f)
&=& D(P_{S \tilde{Z}} \| P_S \times P_{\tilde{Z}} ) \\
&\le& D(P_{KIS\tilde{Z}} \| P_S \times P_{KI\tilde{Z}} ) \\
&=& \sum_{k,i} \frac{1}{|{\cal K}| |{\cal I}|} 
	D(P_{S\tilde{Z}|KI}(\cdot,\cdot|k,i) \| P_S \times P_{\tilde{Z}|KI}(\cdot|k,i)) \\
&=& \sum_{k,i,s} \frac{1}{|{\cal K}| |{\cal I}| |{\cal S}|}
	D(P_{\tilde{Z}|KIS}(\cdot|k,i,s) \| P_{\tilde{Z}|KI}(\cdot|k,i)). 
\end{eqnarray*}
For each $(k,i)$, we use the relation
\begin{eqnarray*}
\lefteqn{
\sum_s \frac{1}{|{\cal S}|} D(P_{\tilde{Z}|KIS}(\cdot|k,i,s) \| P_{\tilde{Z}|KI}(\cdot|k,i)) } \\
&&	+ D(P_{\tilde{Z}|KI}(\cdot|k,i) \| P_{Z|UX_2}(\cdot| u_{ki},x_{2k})) \\
&=& \sum_s \frac{1}{|{\cal S}|} D(P_{\tilde{Z}|KIS}(\cdot|k,i,s) \| P_{Z|UX_2}(\cdot| u_{ki},x_{2k})).
\end{eqnarray*}
By using Lemma \ref{lemma:superposition-resolvability} for 
input distributions $P_{V|UX_2}(\cdot|u_{ki},x_{2k})$ and
$P_{X_1|V}$ and channel $P_{Z|X_1X_2}$, we have
\begin{eqnarray*}
\lefteqn{ \mathbb{E}_{{\cal C}_2 {\cal C}_3} \left[ D(f) \right]} \\
&\le& \sum_{k,i} \frac{1}{|{\cal K}| |{\cal I}|} \\
&&\left[
\frac{1}{\theta |{\cal A}|^{\theta}} e^{\psi(\theta| P_{Z|X_1X_2}(\cdot|\cdot,x_{2k}), P_{X_1|V}, P_{V|UX_2}(\cdot|u_{ki},x_{2k}))} \right. \\
&& + \left. \frac{1}{\theta^\prime |{\cal J}|^{\theta^\prime}} 
	e^{\psi(\theta^\prime| P_{Z|UVX_2}(\cdot|\cdot,u_{ki},x_{2k}), P_{V|UX_2}(\cdot| u_{ki},x_{2k}))} \right] .
\end{eqnarray*}
By taking the average over ${\cal C}_0$ and ${\cal C}_1$, and by noting 
\begin{eqnarray*}
\mathbb{E}_{{\cal C}_0 {\cal C}_1} \left[ \sum_{k,i} \frac{1}{|{\cal K}| |{\cal I}|} \bol{1}[u_{ki} = u, x_{2k} = x_2 ] \right]
	= P_{UX_2}(u,x_2),
\end{eqnarray*}
we have
\begin{eqnarray*}
\lefteqn{ \mathbb{E}_{{\cal C}_0 {\cal C}_1 {\cal C}_2 {\cal C}_3}\left[ D(f) \right]} \\
&\le& \sum_{u,x_2} P_{UX_2}(u,x_2) \\
&&  \left[
\frac{1}{\theta |{\cal A}|^{\theta}} e^{\psi(\theta| P_{Z|X_1X_2}(\cdot|\cdot,x_{2k}), P_{X_1|V}, P_{V|UX_2}(\cdot|u_{ki},x_{2k}))} \right. \\
&& +\left.  \frac{1}{\theta^\prime |{\cal J}|^{\theta^\prime}} 
	e^{\psi(\theta^\prime| P_{Z|UVX_2}(\cdot|\cdot,u_{ki},x_{2k}), P_{V|UX_2}(\cdot| u_{ki},x_{2k}))} \right] \\
&\le& \frac{1}{ \theta |{\cal A}|^\theta} e^{\psi(\theta|P_{Z|X_1 X_2}, P_{X_1|V},P_{UVX_2})} \\
&&	+ \frac{1}{\theta^\prime |{\cal J}|^{\theta^\prime}} e^{\psi(\theta^\prime|P_{Z|UVX_2}, P_{V|UX_2},P_{UX_2})}.
\end{eqnarray*}

\subsection{Proof of Lemma \ref{lemma-fourier-motzkin}}
\label{proof-lemma-fourier-motzkin}

By using the Fourier-Motzkin elimination, we can 
show that $(R_d,R_0,R_1,R_s) \in \tilde{{\cal R}}^{(in)}$
if and only if 
\begin{eqnarray}
R_0 &\le& I(U,X_2; Z), \nonumber \\
R_0 + R_s &\le& I(V;Y|U,X_2) - I(V;Z|U,X_2) \nonumber \\
	&& + I(U,X_2;Y), \label{eq:redundant-one} \\
R_1 + R_s &\le& I(U,V; Y|X_2), \nonumber \\
R_0 + R_1 + R_s &\le& I(V;Y|U,X_2) \nonumber \\
&& + \min[I(U,X_2; Y), I(U,X_2;Z)], \nonumber \\
R_s &\le& I(V; Y|U,X_2) - I(V;Z|U,X_2), \label{eq:security-constraint} \\
R_d + R_1 &\ge& I(X_1; Z|U,X_2), \nonumber \\
R_d &\ge& I(X_1; Z|U,V,X_2) \nonumber 
\end{eqnarray}
are satisfied.
By adding the inequality
\begin{eqnarray*}
R_0 \le I(U,X_2; Y),
\end{eqnarray*}
this inequality and Eq.~(\ref{eq:security-constraint}) imply that
Eq.~(\ref{eq:redundant-one}) is redundant.
Thus, we have ${\cal R}^* \subset \tilde{{\cal R}}^{(in)}$. \qed




\end{document}